# Tailoring the plasticity of topologically close-packed phases via the crystals' fundamental building blocks

Wei Luo [a], Zhuocheng Xie [a,*], Siyuan Zhang [b,*], Julien Guénolé [c,d], Pei-Ling Sun [a], Arno Meingast [e], Amel Alhassan [f], Xuyang Zhou [b], Frank Stein [b], Laurent Pizzagalli [g], Benjamin Berkels [f], Christina Scheu [b], Sandra Korte-Kerzel [a,*]

[a] Institute for Physical Metallurgy and Materials Physics, RWTH Aachen University, Kopernikusstraße 14, 52074 Aachen, Germany

[b] Max-Planck-Institut für Eisenforschung GmbH, Max-Planck-Straße 1, 40237 Düsseldorf, Germany

[c] Université de Lorraine, CNRS, Arts et Métiers ParisTech, LEM3, 57070 Metz, France

[d] Labex Damas, Université de Lorraine, 57070 Metz, France

[e] Thermo Fisher Scientific, 5651 GH De Schakel 2, Eindhoven, Netherlands

[f] Institute for Advanced Study in Computational Engineering Science, RWTH Aachen University, Schinkelstr. 2, 52062 Aachen, Germany

[g] Institut Pprime, CNRS UPR 3346, Université de Poitiers, SP2MI, Boulevard Marie et Pierre Curie, TSA 41123, 86073 Poitiers Cedex 9, France

*Corresponding authors: Xie@imm.rwth-aachen.de (Zhuocheng Xie); siyuan.zhang@mpie.de (Siyuan Zhang); Korte-Kerzel@imm.rwth-aachen.de (Sandra Korte-Kerzel)

*Keywords: plasticity; intermetallics; dislocations; HR-TEM; DFT*

## Abstract

Brittle topologically close-packed precipitates form in many advanced alloys. Due to their complex structures little is known about their plasticity. Here, we present a strategy to understand and tailor the deformability of these complex phases by considering the Nb-Co μ-phase as an archetypal material. The plasticity of the Nb-Co μ-phase is controlled by the Laves phase building block that forms parts of its unit cell. We find that between the bulk $C15-NbCo_2$ Laves and Nb-Co μ-phase, the interplanar spacing and local elastic modulus of the Laves phase building block change, leading to a strong reduction in hardness and elastic modulus, as well as a transition from synchroshear to crystallographic slip. Furthermore, as the composition changes from $Nb_6Co_7$ to $Nb_7Co_6$, the Co atoms in the triple layer are substituted such that the triple layer of the Laves phase building block becomes a slab of pure Nb, resulting in inhomogeneous changes in elasticity and a transition from crystallographic slip to a glide-and-shuffle mechanism. These findings open opportunities to purposefully tailor the plasticity of these topologically close-packed phases in bulk, but at the atomic scale of interplanar spacing and local shear modulus of the fundamental crystal building blocks in their large unit cells.

**Introduction**

New materials with superior strength, ductility and high temperature capability are at the heart of future developments to meet challenges in mobility, energy conversion and sustainability. In knowledge-based development of structural materials, we commonly design and manipulate the materials' internal (micro)structures at several length scales to control the physical mechanisms governing deformation. This strategy has been particularly successful in the invention of advanced high-strength steels outperforming those used over previous decades within years by an order of magnitude in strength and deformability [1]. While the origin of mechanical strength, toughness and creep resistance is well studied in metals, very little is known about the properties of intermetallics, despite their wide use as reinforcement phases and their sheer number and variability [2]. If we were to understand the physical origin of their mechanical properties based on the underlying deformation mechanisms as well as we do now in metals and metallic alloys, we could manipulate their strength and toughness. Predicting and controlling the plasticity of intermetallics is of great interest as we could tailor highly alloyed materials to form only those intermetallics that reinforce but do not introduce damage by brittle cracking.

A large fraction of complex intermetallics can be simplified as an intergrowth of a few simple fundamental units [3]. By considering the simple fundamental building blocks of complex intermetallics, rather than entire large unit cells, it provides a pathway to unravel the plasticity of complex intermetallics based on the deformation mechanisms of the simple fundamental building blocks that combine to form the complex structures, and the relationships between the crystal structure and chemical composition of the complex intermetallics and the mechanical behavior of the simple fundamental building blocks. Furthermore, the recurrent nature of the few fundamental building blocks [3] will allow a transfer of knowledge to a large number of complex phases. Such knowledge would allow us to mine the existing, enormous crystal databases for exceptional phases that could form the basis for future high performance alloys.

The topologically close-packed (TCP) phases are commonly found in superalloys [4] but their plasticity is lesser-known due to their complex crystal structures. Laves phases are the most common intermetallic compounds with a relatively simple TCP structure formed by stacking of quadruple Laves phase layers which are composed of a kagomé layer and a triple layer [5]. The Laves phases have been confirmed to plastically deform via a synchroshear mechanism [6], which consists of two shears in different directions on adjacent atomic planes of the triple layer structural unit. Although synchroshear is geometrically and energetically more favorable than the common crystallographic slip in the Laves phase [7], the motion of the associated synchro-Shockley partial dislocations needs to overcome high energy barriers with the help of thermal activation [8], leading to the intrinsically high strength and brittleness of the Laves phases. The μ-phase is a common intermetallic precipitate phase with a complex TCP structure in superalloys [4a, 9] but our knowledge about its plasticity is still limited. As the μ-phase consists of the $MgCu_2$ Laves phase layers interspersed by monoatomic layers of $Zr_4Al_3$ [10] (Fig. 1), synchroshear is therefore thought to also occur in the μ-phase [11]. However, the local atomic environment of the Laves phase building block in the μ-phase is different from that in the Laves phases, which may lead to different interplanar spacing, different local stiffness of the interatomic bonds, and thus distinct mechanical behavior. Moreover, the μ-phase can have a wide composition range and it can be stable at off-stoichiometric compositions where the atomic configurations of the Laves phase building block is substantially modified by the constitutional defects [12]. Therefore, we use the μ-phase as a model to explore the influence of crystal structure and composition on the mechanical behavior of the fundamental Laves phase building block to understand the plasticity of complex μ-phase, and open up opportunities for controlling the deformation mechanisms and tailoring the mechanical properties of complex TCP phases.

In the present work, we investigated the mechanical properties of the Nb-Co μ-phase and the closely related $C15$-$NbCo_2$ Laves phase by nanoindentation tests and found that a reduction in both hardness and elastic modulus can be achieved by changing the structure and composition. Furthermore, we revealed how the structure and composition affect the mechanical properties and deformation mechanisms of the Laves phase building block by means of aberration-corrected high-angle annular dark-field scanning transmission electron microscopy (HAADF-STEM) combined with atomic-resolution energy dispersive X-ray spectroscopy (EDS) analysis and density functional theory (DFT) calculations.

**Results and discussion**

**Influence of composition on the atomic configurations.**

In the μ-phase, while the large atoms prefer to occupy the positions with high coordination numbers (CN = 14, 15 and 16), the small atoms tend to occupy the positions with a low coordination number (CN = 12) [10]. The stoichiometric composition of the μ-$Nb_6Co_7$ phase is about 46.2 at.% Nb. As the composition deviates from the stoichiometric composition, constitutional defects, such as anti-site atoms, are expected in the Nb-Co μ-phase [12]. In order to study the atomic configurations of the Nb-Co μ-phase at off-stoichiometric compositions, the elemental maps of the $Nb_{6.4}Co_{6.6}$ (Fig. 1 (a)) and $Nb_7Co_6$ (Fig. 1 (b)) alloys (Supplementary Table 1) of the Nb-Co μ-phase were measured by atomic-resolution EDS analysis. As shown in Fig. 1 (a) and (b), the positions with high coordination numbers (CN = 14, 15 and 16) are exclusively occupied by large Nb atoms, and the positions in the kagomé layers which have a low coordination number (CN = 12) are fully occupied by small Co atoms. However, in the middle (CN = 12) of the triple layers, Nb was detected in both the $Nb_{6.4}Co_{6.6}$ and $Nb_7Co_6$ alloys. While both Co and Nb occur in the middle (CN = 12) of the triple layers in the $Nb_{6.4}Co_{6.6}$ alloy (Fig. 1 (a)), hardly any Co remains in the middle (CN = 12) of the triple layers in the $Nb_7Co_6$ alloy (Fig. 1 (b)). This indicates that, in the Nb-rich Nb-Co μ-phase, the excess Nb atoms substitute the Co atoms specifically in the middle of the triple layers. Therefore, as the composition is reversed from $Nb_6Co_7$ to $Nb_7Co_6$, the general structure of the Nb-Co μ-phase (Fig. 1 (c)) is preserved but the Co atoms in the triple layers are replaced by Nb atoms, and thus the puckered close-packed Nb-Co-Nb triple layer of the Laves phase building block become a slab of pure Nb (Fig. 1 (d)).

The interplanar spacing between the basal planes of the middle and the bottom of the triple layer $d_t$, and the interplanar spacing between the basal planes of the bottom of the triple layer and kagomé layer $d_{t-k}$ (Fig. 1 (c)), measured from HAADF-STEM images of the $Nb_{6.4}Co_{6.6}$ alloy are $0.32 \pm 0.02$ Å and $1.70 \pm 0.12$ Å, respectively. The respective values of $d_t$ and $d_{t-k}$ of the $Nb_7Co_6$ alloy are $0.43 \pm 0.02$ Å and $1.67 \pm 0.08$ Å. These measurements were made using a novel mathematical analysis on STEM images based on variational methods for motif extraction [13]. We also investigated the interplanar spacings of the Nb-Co μ-phase and the closely related C15-NbCo$_2$ Laves phase by DFT calculations. As the crystal structure changes from C15-NbCo$_2$ to μ-Nb$_6$Co$_7$, $d_t$ decreases dramatically from 0.481 Å to 0.272 Å (Supplementary Table 2) whereas $d_{t-k}$ increases from 1.454 Å to 1.700 Å. As the composition of the Nb-Co μ-phase changes from $Nb_6Co_7$ to $Nb_7Co_6$, $d_t$ changes from 0.272 Å to 0.363 Å, while $d_{t-k}$ remains nearly constant. These DFT calculations are consistent with the results measured from experimental HAADF-STEM images.

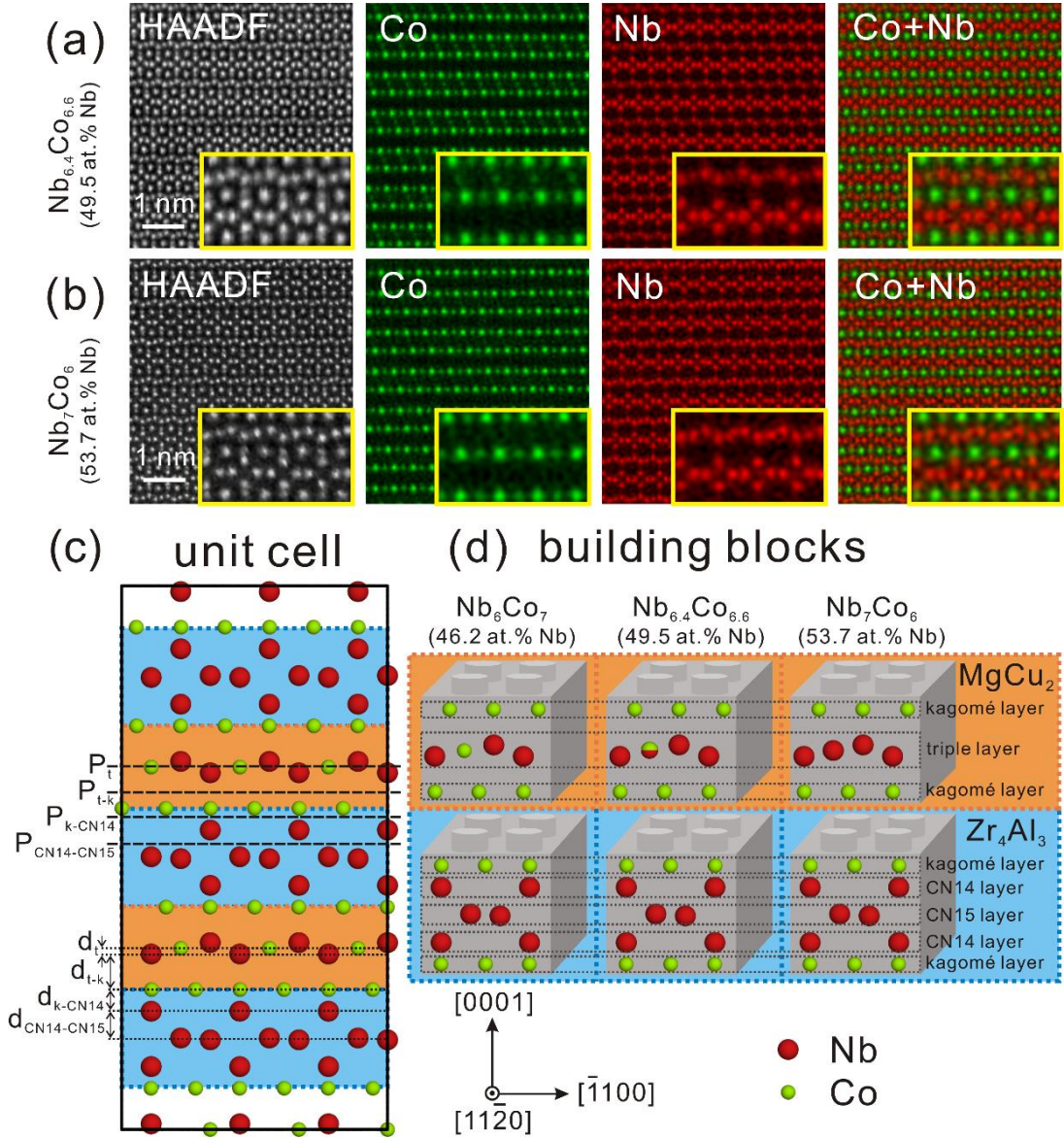

*Fig. 1 HAADF-STEM images and the corresponding atomic resolution EDS maps of the (a) Nb$_{6.4}$Co$_{6.6}$ and (b) Nb$_7$Co$_6$ alloys showing the distributions of Co and Nb in the Nb-Co μ-phase. Schematics of the $[11\bar{2}0]$ projection of the Nb-Co μ-phase showing the (c) unit cell of μ-Nb$_6$Co$_7$ and (d) the stacking of the MgCu$_2$ Laves phase and Zr$_4$Al$_3$ building blocks in μ-Nb$_6$Co$_7$, μ-Nb$_{6.4}$Co$_{6.6}$ and μ-Nb$_7$Co$_6$. P$_t$, P$_{t-k}$, P$_{k-CN14}$, and P$_{CN14-CN15}$ denote the basal plane between the middle and the bottom of the triple layer, the basal plane between the bottom of the triple layer and the kagomé layer, the basal plane between the kagomé layer and the CN14 layer, and the basal plane between the CN14 layer and the CN15 layer, respectively. d$_t$, d$_{t-k}$, d$_{k-CN14}$, and d$_{CN14-CN15}$ are the corresponding interplanar spacings.*

**Influence of composition on the mechanical response.**

Nanoindentation tests were performed on the $Nb_{6.4}Co_{6.6}$, $Nb_{6.9}Co_{6.1}$ and $Nb_7Co_6$ alloys to study the influence of composition on the hardness and elastic modulus of the Nb-Co μ-phase. As shown in the *post-mortem* scanning electron microscopy (SEM) images (Figs. 2 (a-c) and Supplementary Figs. 1 (a-d)), the straight basal slip traces around the indentations reveal that plastic deformation of the Nb-Co μ-phase occurs in an anisotropic manner, predominantly by basal slip. In contrast to the hexagonal Laves phase, which can deform plastically by both basal and non-basal slip at room temperature [14], the Nb-Co μ-phase only shows visible slip traces of basal slip even though it is in an orientation where non-basal slip has a high Schmid factor.

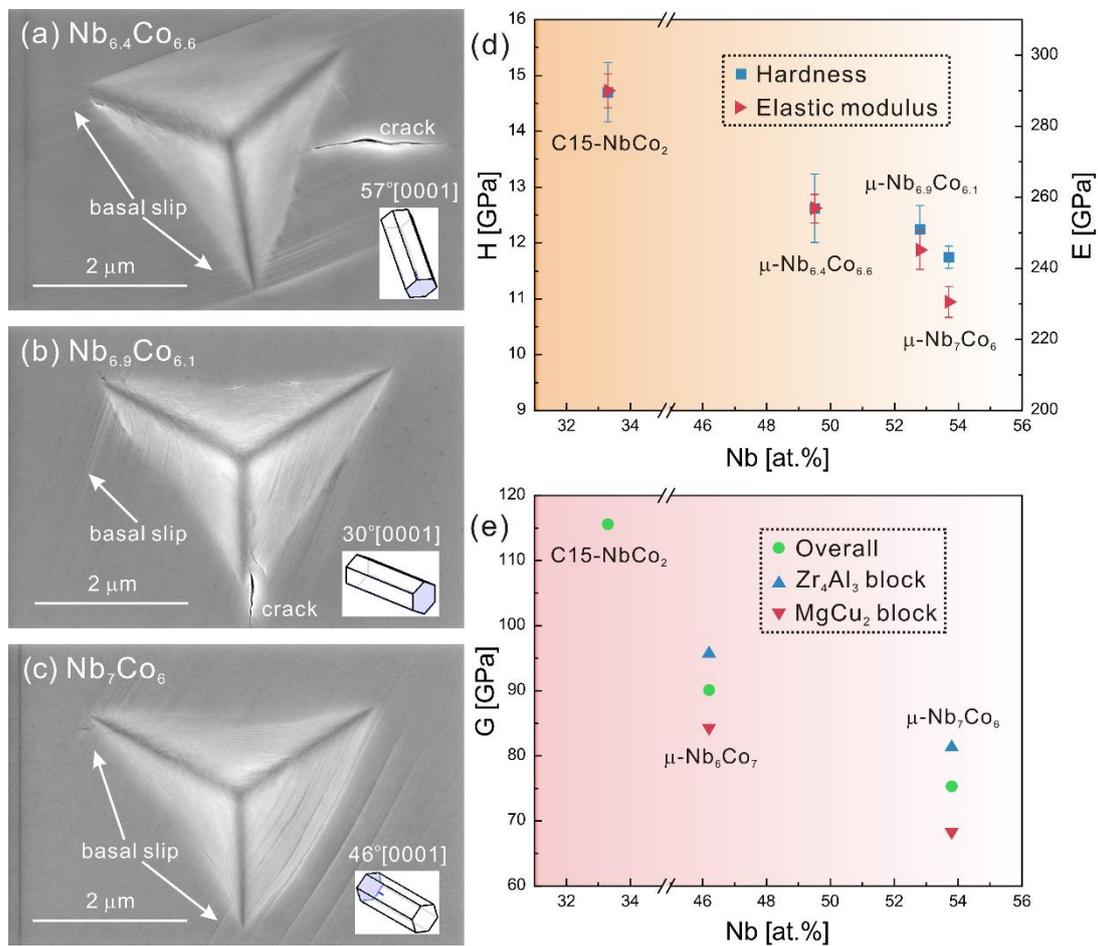

*Fig. 2 Representative SEM images of indentations of the (a) $Nb_{6.4}Co_{6.6}$, (b) $Nb_{6.9}Co_{6.1}$ and (c) $Nb_7Co_6$ alloys showing the surface slip traces after indentation tests. The orientations are denoted by the angle between the loading axis and the c-axis of the Nb-Co μ-phase. Unit cells representing the orientations are inserted. (d) The composition dependence of hardness and elastic modulus of the Nb-Co μ-phase measured by nanoindentation tests. (e) Calculated basal plane shear moduli of the overall crystal and the $Zr_4Al_3$ and $MgCu_2$ building blocks of C15-$NbCo_2$, μ-$Nb_6Co_7$ and μ-$Nb_7Co_6$.*

Nanoindentation tests performed in selected grains with different orientations (Supplementary Figs. 2 (a-i)) reveal that the influence of orientation on the hardness and elastic modulus of the Nb-Co μ-phase is negligible. The hardness (Fig. 2 (d)) of the Nb-Co μ-phase is lower than that of the C15-NbCo$_2$ Laves phase. As the Nb content increases, the hardness of the Nb-Co μ-phase decreases from $12.6 \pm 0.6$ GPa at 49.5 at.% Nb to $11.9 \pm 0.3$ GPa at 52.8 at.% Nb, and to $10.9 \pm 0.3$ GPa at 53.7 at.% Nb.

The elastic modulus (Fig. 2 (d)) measured by nanoindentation tests shows a similar compositional trend as the hardness. The measured elastic modulus (Fig. 2 (d)) of the Nb-Co μ-phase is lower than that of the C15-NbCo$_2$ Laves phase, and it decreases from $257 \pm 4$ GPa at 49.5 at.% Nb to $251 \pm 7$ GPa at 52.8 at.% Nb, and to $243 \pm 3$ GPa at 53.7 at.% Nb. The elastic properties and elastic anisotropy of μ-Nb$_6$Co$_7$ and μ-Nb$_7$Co$_6$ were also investigated by DFT calculations. The overall Young's modulus and shear modulus of μ-Nb$_7$Co$_6$ are lower than those of μ-Nb$_6$Co$_7$ (Supplementary Table 3), which is consistent with the compositional trend of elastic modulus measured from the nanoindentation tests. The overall basal plane shear modulus and the respective basal plane shear moduli of the Zr$_4$Al$_3$ and MgCu$_2$ building blocks of the Nb-Co μ-phase were calculated and compared with the $\langle 111 \rangle$ plane shear modulus of the closely related C15-NbCo$_2$ Laves phase in Fig. 2 (e). The overall basal plane shear modulus of the Nb-Co μ-phase is significantly lower than the $\langle 111 \rangle$ plane shear modulus of the C15-NbCo$_2$ Laves phase, and it decreases from 82.6 to 72.4 GPa as the composition changes from Nb$_6$Co$_7$ to Nb$_7$Co$_6$. Moreover, although the shear modulus is isotropic on the basal plane of the Nb-Co μ-phase, the basal plane shear moduli of the two building blocks are different. As shown in Fig. 2 (e), the basal plane shear modulus of the MgCu$_2$ building block is lower than that of the Zr$_4$Al$_3$ building block, indicating that the MgCu$_2$ building block is more compliant, and thus is more prone to shear straining than the Zr$_4$Al$_3$ building block in the Nb-Co μ-phase. The basal plane shear modulus of the MgCu$_2$ building block is about 12 % and 16 % lower than that of the Zr$_4$Al$_3$ building block in μ-Nb$_6$Co$_7$ and μ-Nb$_7$Co$_6$, respectively. This suggests that as the composition changes from Nb$_6$Co$_7$ to Nb$_7$Co$_6$, the elastic inhomogeneity between the Zr$_4$Al$_3$ and MgCu$_2$ building blocks becomes more pronounced.

**Dislocation structures in Nb$_{6.4}$Co$_{6.6}$ and Nb$_7$Co$_6$ alloys.**

As shown in the *post-mortem* SEM image (Supplementary Fig. 1 (b)), the Nb$_{6.4}$Co$_{6.6}$ alloy shows a high density of straight basal slip traces around the indentation when indented along 83º [0001] direction. In order to study the dislocations structures, a TEM lamella of the 83º [0001] indentation of the Nb$_{6.4}$Co$_{6.6}$ alloy was prepared. The bright-field TEM image taken with $g = \bar{1}101$ (Supplementary Fig. 3 (a)) shows that there are numerous dislocations and several widely extended stacking faults parallel to the basal plane underneath the indentation. When imaged with $g = 0003$ (Supplementary Fig. 3 (b)), the defects are out of contrast, which confirms that their Burgers vectors are on the basal plane. As the wide stacking faults start at the top surface and extend to the bottom of the lamella, they are assumed to be grown-in defects. While the contrast of the widely extended stacking faults is well visible and their widths change when the sample is tilted, no contrast of stacking faults between the dislocations can be observed. The dislocations are therefore either full dislocations without dissociation or the dissociation distance is too narrow to be resolved in the bright-field TEM images.

In order to further resolve the structures of the dislocations, we performed aberration-corrected HAADF-STEM imaging. A full dislocation in the $Nb_{6.4}Co_{6.6}$ alloy is shown in Fig. 3 (a). The closure failure of the Burgers circuit reveals that the imaged core corresponds to a full dislocation with $b = \frac{1}{3}\langle \bar{2}110 \rangle$ lying between the kagomé layer and the triple layer of the $MgCu_2$ Laves phase building block. While most of the defects imaged in the plastic zone of the indentation were confirmed to be full dislocations with $b = \frac{1}{3}\langle \bar{2}110 \rangle$, a few partial dislocations bounding stacking faults on the basal plane were also observed. The partial dislocation core and the associated stacking fault in the $Nb_{6.4}Co_{6.6}$ alloy are shown in Fig. 3 (b). The stacking fault is located in the triple layer of the $MgCu_2$ Laves phase building block and the stacking sequence of the triple layer changes to the twinned variant in the stacking fault. The closure failure of the Burgers circuit confirms that the partial dislocation has a Burgers vector of $b = 1/3\langle 0\bar{1}10 \rangle$.

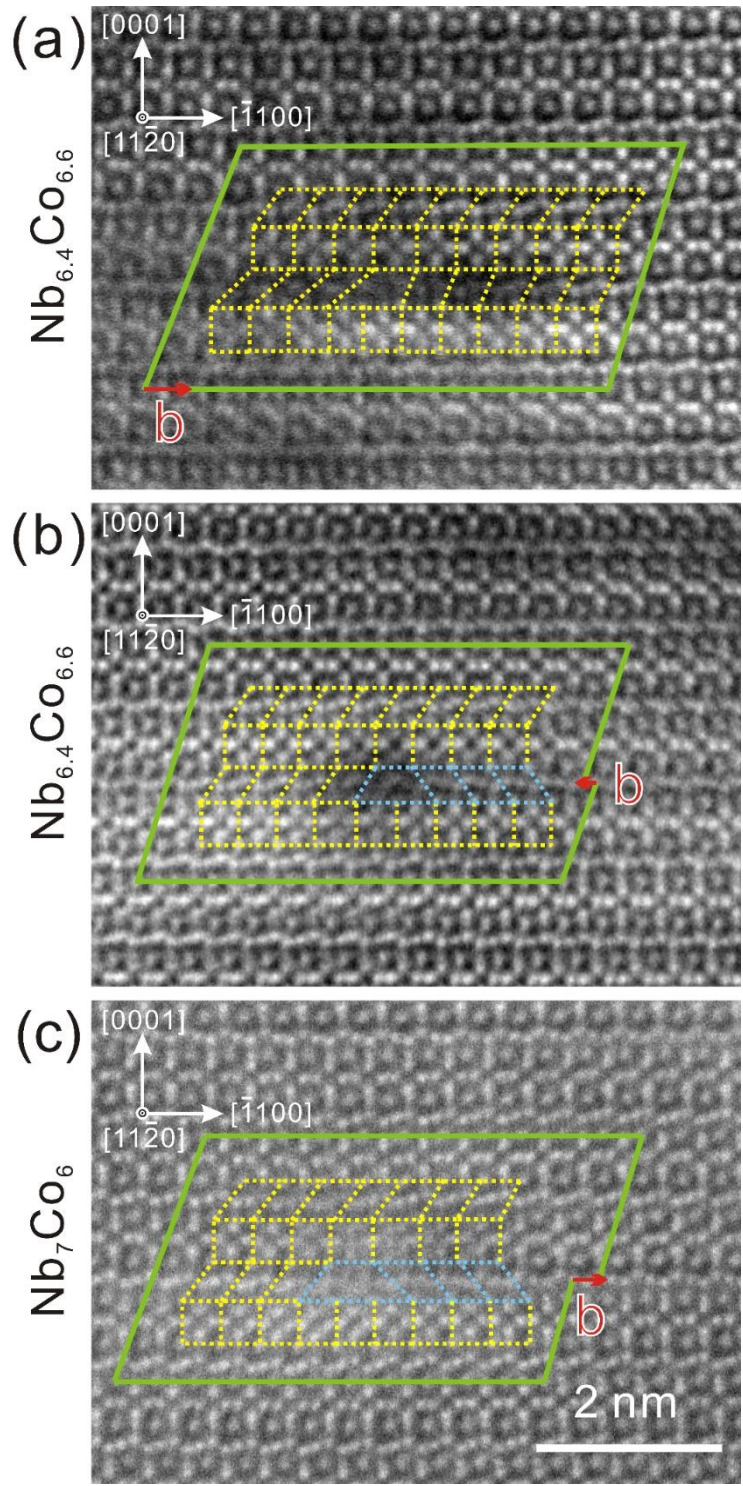

*Fig. 3 Atomic configurations of the defects on the basal plane of the Nb-Co μ-phase under an indentation viewed along* $[11\bar{2}0]$ *axis. HAADF-STEM images of (a) a full dislocation in the $Nb_{6.4}Co_{6.6}$ alloy viewed end-on, (b) a partial dislocation bounding a stacking fault in the $Nb_{6.4}Co_{6.6}$ alloy, and (c) a partial dislocation bounding a stacking fault in the $Nb_7Co_6$ alloy. The Burgers vectors marked by red arrows are measured from the closure failures of the green Burgers circuits. The stacking of the building blocks of the Nb-Co μ-phase and the stacking faults are outlined using yellow and blue tiles, respectively.*

In the $Nb_7Co_6$ alloy, a high density of straight basal slip traces was observed around the indentation with a loading axis of 46º [0001] (Fig. 2 (c)). In the bright-field TEM image from such an indentation, numerous planar defects parallel to the basal plane and bounded by partial dislocations were observed using a two-beam condition with $g = 0\bar{1}11$ (Supplementary Fig. 3 (c)). The planar defects are out of contrast for $g = 0003$ (Supplementary Fig. 3 (d)), which confirms that the Burgers vectors of the planar defects in $Nb_7Co_6$ alloy are also on the basal plane. The structure of a stacking fault bound by partial dislocations in this sample was also imaged by HAADF-STEM (Fig. 3 (c)) along the $[11\bar{2}0]$ direction. At the dislocation core, the parallelograms containing triple layers change direction, indicating that the stacking fault is located in the $MgCu_2$ Laves phase building block. The closure failure of the Burgers circuit around the dislocation core reveals the partial dislocation Burgers vector $b = \frac{1}{3}[\bar{1}100]$.

**Energetically favorable slip events**

In order to reveal the energy paths of crystallographic slip in the Nb-Co μ-phase, we calculated the generalized stacking fault energy (GSFE) curves of the crystallographic slip on different basal planes along the $\frac{1}{3}\langle 11\bar{2}0\rangle$ direction in μ-$Nb_6Co_7$ (Fig. 4 (a)) and μ-$Nb_7Co_6$ (Fig. 4 (b)). The crystallographic slip mechanism [7a, 7c] involves the sliding of one half of a crystal relative to the other half of the crystal on the basal plane along a crystallographic direction perpendicular to the stacking direction. Among the four possible sets of basal planes $P_{CN14-CN15}$, $P_{k-CN14}$, $P_{t-k}$, and $P_t$, we find that crystallographic slip on $P_{t-k}$ exhibits the lowest energy barrier in both μ-$Nb_6Co_7$ (Fig. 4 (a)) and μ-$Nb_7Co_6$ (Fig. 4 (b)). The GSFE curves of crystallographic slip on $P_{t-k}$ along the $\frac{1}{3}\langle\bar{1}100\rangle$ direction in μ-$Nb_6Co_7$ and μ-$Nb_7Co_6$ are given in Fig. 4 (c). The GSFE curve of crystallographic slip on $P_{t-k}$ along the $\frac{1}{3}\langle\bar{1}100\rangle$ direction in μ-$Nb_7Co_6$ shows a lower energy barrier and a more pronounced local minimum than that in μ-$Nb_6Co_7$. Crystallographic slip on $P_{t-k}$ therefore becomes easier and dislocation dissociation is more likely to occur as the composition of the Nb-Co μ-phase changes from $Nb_6Co_7$ to $Nb_7Co_6$. Moreover, the Nb-Co μ-phase has a lower energy barrier to basal slip on $P_{t-k}$ than the C15-$NbCo_2$ Laves phase (Supplementary Fig. 4 (a)), presumably due to the larger interplanar spacing $d_{t-k}$ in the Nb-Co μ-phase. Therefore, crystallographic slip on $P_{t-k}$ becomes increasingly more favorable as the Nb-Co TCP phase changes from the Laves C15-$NbCo_2$ to μ-$Nb_6Co_7$, and to μ-$Nb_7Co_6$.

In order to explore the minimum energy paths (MEPs) and the associated atomic mechanisms of the basal slip mechanisms other than the classical crystallographic slip in the Nb-Co μ-phase, DFT nudged elastic band (NEB) calculations were performed for basal slip on $P_t$ and $P_{t-k}$ of the $MgCu_2$ Laves phase building block. To ensure a constant strain state throughout the NEB calculations, the initial configuration was elastically strained with a shear of a full Burgers vector, which is equivalent to the plastic strain of the final configuration (Supplementary Fig. 5). The NEB calculations predict that synchroshear is the energetically most favorable slip mechanism on $P_t$ in both μ-$Nb_6Co_7$ and μ-$Nb_7Co_6$ (Fig. 4 (d)). The NEB calculations were also performed on $P_t$ of the C15-$NbCo_2$ Laves phase for comparison. The computed MEP and the atomic mechanism of basal slip on the $P_t$ of the C15-$NbCo_2$ Laves phase are also associated with the synchroshear mechanism (Supplementary Figs. 4 (b) and (c)), which agrees with previous *ab-initio* [7a] and atomistic simulations [7c, 8a]. The energy barrier to synchroshear in the Nb-Co μ-phase (Fig. 4 (d)) is much higher than that in the C15-$NbCo_2$ Laves phase (Supplementary Fig. 4 (b)) presumably due to the reduced interplanar spacing $d_t$, and the energy barrier to synchroshear in the Nb-Co μ-phase increases dramatically from 1999.7 to 2677.8 mJ/m$^2$ as the composition changes from $Nb_6Co_7$ to $Nb_7Co_6$.

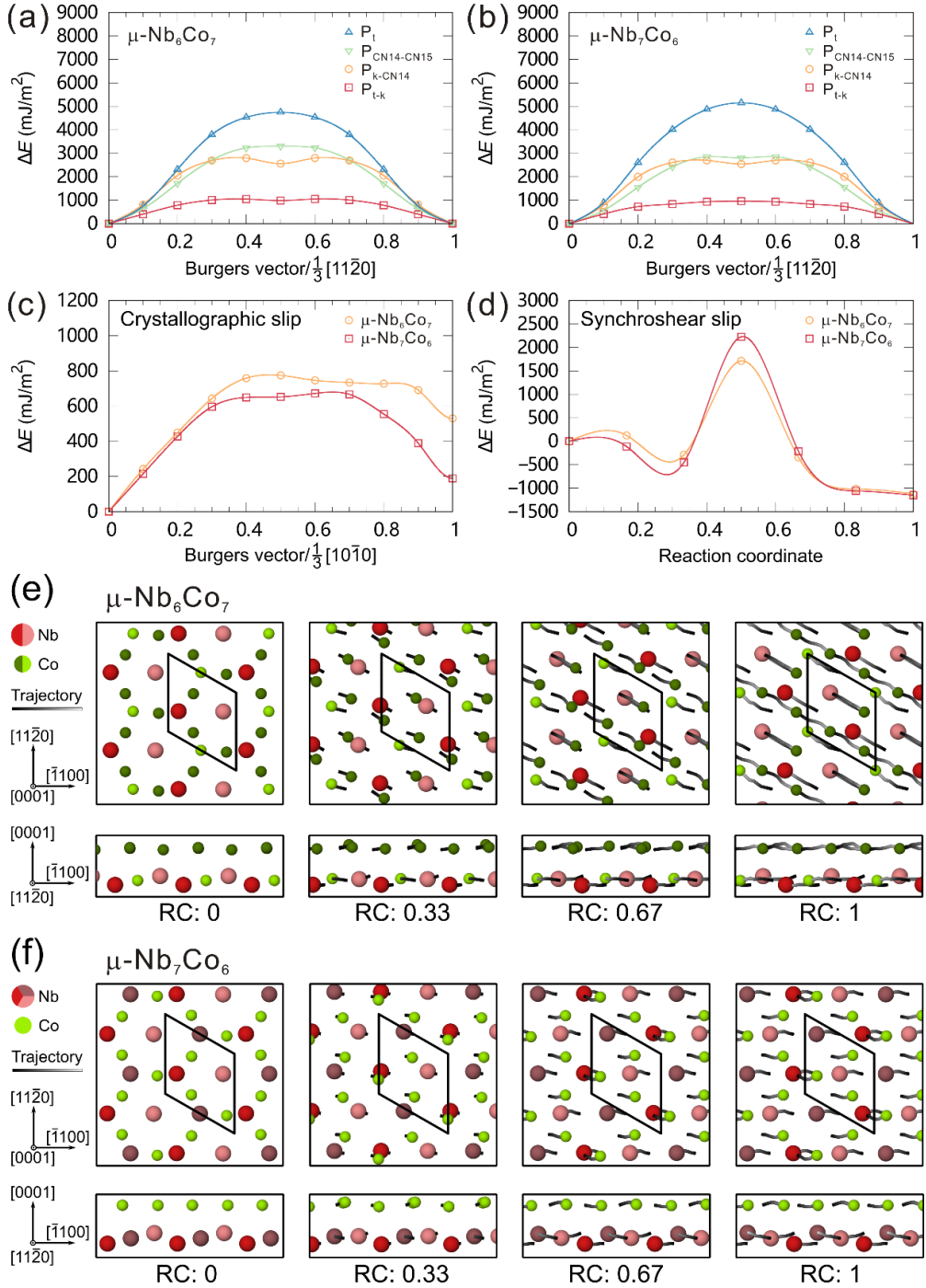

Fig. 4 Assessment of the slip mechanisms in μ-Nb₆Co₇ and μ-Nb₇Co₆ using DFT. GSFE curves of the crystallographic slip along $\frac{1}{3}[11\bar{2}0]$ direction on different basal planes in (a) μ-Nb₆Co₇ and (b) μ-Nb₇Co₆. Energy profiles of (c) the crystallographic slip along $\frac{1}{3}[10\bar{1}0]$ direction on P$_{t-k}$ calculated by GSFE and (d) synchroshear calculated using the NEB method. Snapshots of basal slip on P$_{t-k}$ via (e) the crystallographic slip mechanism in μ-Nb₆Co₇ and via (f) the glide and shuffle mechanism in μ-Nb₇Co₆. Large (colored in different shades of red) and small (colored in green) atoms are Nb and Co atoms, respectively. Trajectories of atoms are colored in a black-grey gradient according to the reaction coordinate.

The synchroshear mechanism is less sensitive to the applied shear strain owing to the thermally activated nature of the kink-pair nucleation and propagation of synchro-Shockley dislocations [8a]. In contrast, the computed MEPs show that the energy barriers to basal slip on $P_{t-k}$ in $\mu$-$Nb_6Co_7$ (Supplementary Fig. 6 (a)) and $\mu$-$Nb_7Co_6$ (Supplementary Fig. 6 (b)) both vanish due to the large pre-strain. According to the NEB calculations, the associated atomic mechanisms of basal slip on $P_{t-k}$ in $\mu$-$Nb_6Co_7$ and $\mu$-$Nb_7Co_6$ are different. The intermediate states of the basal slip on $P_{t-k}$ in $\mu$-$Nb_6Co_7$ (Fig. 4 (e)) show that basal slip on $P_{t-k}$ in $\mu$-$Nb_6Co_7$ occurs via a crystallographic slip mechanism along the $\frac{1}{3}\langle 11\bar{2}0\rangle$ direction. The triple layer moves as a whole along the

$\frac{1}{3}\langle 11\bar{2}0\rangle$ direction relative to the kagomé layer above. The stacking sequence of the triple layer in $\mu$-$Nb_6Co_7$ therefore remains unchanged during the glide process. However, the intermediate states of basal slip on $P_{t-k}$ in $\mu$-$Nb_7Co_6$ (Fig. 4 (f)) show that basal slip on $P_{t-k}$ in $\mu$-$Nb_7Co_6$ occurs via a glide and shuffle mechanism. An out-of-plane atomic shuffling between the top and the middle of the triple layer occurs during the glide process on the $P_{t-k}$ plane along $\langle \bar{1}100\rangle$ direction. The Nb atoms on the top move to the middle of the triple layer, and meanwhile the Nb atoms in the middle adjust their positions to the top of the triple layer. After the glide and shuffle process, the triple layer changes to the twinned variant, and a stable stacking fault with a low stacking fault energy of 7.6 mJ/m$^2$ is created. The configuration of the stacking fault is consistent with those widely observed in the HAADF-STEM images of $\mu$-$Nb_7Co_6$ (Fig. 3 (c)). Although in $\mu$-$Nb_7Co_6$ the configuration and the associated stacking fault energy of this type of stacking fault are the same as the stacking fault created by synchroshear, the stacking faults in the $Nb_7Co_6$ alloy are more likely to be created by the glide and shuffle mechanism rather than the synchroshear mechanism due to the high energy barrier to synchroshear. However, this type of stacking fault created by the glide and shuffle mechanism is unlikely to occur in C15-$NbCo_2$ and $\mu$-$Nb_6Co_7$ since it disrupts the stable Nb-Co-Nb configuration of the triple layer (Supplementary Fig. 7), and thus the energy of this type of stacking fault is very high in C15-$NbCo_2$ (802.1 mJ/m$^2$) and $\mu$-$Nb_6Co_7$ (471.7 mJ/m$^2$) (Supplementary Table 4). Although the $Nb_{6.4}Co_{6.6}$ alloy mainly deforms by full dislocation slip on $P_{t-k}$, given that nearly 43.5 % of the Co atoms in the middle of the triple layers are replaced by Nb atoms at 49.5 at.% Nb, the energy of the stacking fault created by the glide and shuffle mechanism might be reduced, and thus the full dislocations might be able to dissociate with the help of local fluctuations in composition.

In summary, we can now interpret the plasticity of Nb-Co $\mu$-phase by elucidating the influences of crystal structure and composition on the mechanical behavior of the Laves phase building block. As the crystal structure changes from the C15-NbCo$_2$ Laves phase to the Nb-Co μ-phase, the interplanar spacing $d_t$ decreases and the interplanar spacing $d_{t-k}$ increases. As a result of the changes in interplanar spacings of the Laves phase building block, the energy barrier to synchroshear dramatically increases, whereas the energy barrier to basal slip between the triple layer and kagomé layer decreases. Thus, the basal slip mechanism tends to change from synchroshear in the C15-NbCo$_2$ to basal slip between the triple layer and kagomé layer in the Nb-Co μ-phase. The excess Nb atoms in the Nb-rich Nb-Co μ-phase are accommodated by the site specific substitutions of the Co atoms in the triple layer of the MgCu$_2$ Laves phase building block. As the composition changes from Nb$_6$Co$_7$ to Nb$_7$Co$_6$, the Nb-Co-Nb triple layer become a slab of pure Nb, resulting in a reduction in elastic modulus, especially the basal plane shear modulus of the MgCu$_2$ Laves phase building block, and an increased elastic inhomogeneity between the Zr$_4$Al$_3$ and MgCu$_2$ Laves phase building blocks. The changes in the atomic configurations of the MgCu$_2$ Laves phase building block also lead to a reduction in energy barrier to basal slip on the P$_{t-k}$ plane and formation of stable stacking faults via the glide and shuffle mechanism. Therefore, as the composition of the Nb-Co μ-phase changes from Nb$_6$Co$_7$ to Nb$_7$Co$_6$, basal slip on the P$_{t-k}$ plane becomes easier and a transition of basal slip mechanism from the crystallographic slip to the glide and shuffle mechanism occurs.

The strong influences of the structure and chemical composition on the mechanical behavior of the Laves phase building block sheds light on how to control the mechanical properties of the complex μ-phase. It therefore opens up opportunities to make the hard and brittle μ-phase phases much softer and deformable at even ambient temperature [15], e.g., through alloying with elements occupying specific sites of the unit cell. Furthermore, these insights also have implications for understanding of the plasticity of complex intermetallic phases with a layered structure in general. The plasticity of the complex intermetallic phases built from layers of recurring building blocks, such as ternary transition metal nitrides/carbides (MAX phases) [16], can be governed by the mechanisms active in the fundamental sub-units and may be tailored by tuning local bonding conditions and lattice spacing. To study the interplay of interplanar spacing, local shear modulus and plasticity in the recurrent fundamental building blocks will also allow us to transfer our understanding of their deformation to a large number of complex intermetallic phases that may be newly considered or discovered for the development of new high performance materials.

## Acknowledgements

The authors thank Dunming Wu, David Beckers and Arndt Ziemons for their help in sample preparation, Martin Heller and Risheng Pei for their help in focused ion beam milling, James Best for the help in nanoindentation and micropillar compression tests, as well as Stefanie Sandlöbes-Haut for fruitful discussions. This project has received funding from the European Research Council (ERC) under the European Union's Horizon 2020 Research and Innovation Programme (Grant Agreement No. 852096 FunBlocks). Funding by the Deutsche Forschungsgemeinschaft (DFG) in project KO 4603/2-2 (project ID 437514011) and SFB1394 Structural and chemical atomic complexity – from defect phase diagrams to material properties (Project ID 409476157) is gratefully acknowledged. Simulations were performed with computing resources granted by RWTH Aachen University under project (p0020267).

**Data availability statement**

The image and simulation data underlying this work will be made available publicly after it has been accepted for publication. The underlying data will be made available on request to the reviewers (via the journal) at any time during the review.